\documentclass[12pt,a4paper,final]{iopart}

\usepackage{iopams}  
\usepackage{graphicx}
\usepackage[breaklinks=true,colorlinks=true,linkcolor=blue,urlcolor=blue,citecolor=blue]{hyperref}
\usepackage{braket}

\newcommand{\tauWg}{\tau_\mathrm{W}^\mathrm{g}}
\newcommand{\D}{\mathrm{d}}
\begin{document}

\title{Comment on ``Time delays in molecular photoionization"}

\author{Denitsa Baykusheva}
\address{Laboratorium f\"ur Physikalische Chemie, ETH Z\"urich,\\
Vladimir-Prelog-Weg 2, 8093 Z\"urich, Switzerland}
\author{Hans Jakob W\"orner}
\address{Laboratorium f\"ur Physikalische Chemie, ETH Z\"urich,\\
Vladimir-Prelog-Weg 2, 8093 Z\"urich, Switzerland}
\ead{woerner@phys.chem.ethz.ch}

\begin{abstract}
In a recent article by P. Hockett \textit{et al.}~\cite{Hockett16}, time delays arising in the context of molecular single-photon ionization are investigated from a theoretical point of view. We argue that one of the central equations derived in the paper is incorrect and present a reformulation that is consistent with the established treatment of angle-dependent scattering delays~\cite{Eisenbud48, Wigner55, Smith60, Nussenzveig72}. 

\end{abstract}

\vspace{2pc}
In their recent article~\cite{Hockett16} investigating the spectral and angular dependence of time delays occurring during the process of molecular single-photon ionization, Hockett \textit{et al.} provide the following general expression (Eq.~(4) in the original publication) for the time delay $\tauWg(k, \theta, \phi)$  associated with the outgoing photoelectron with momentum $k$ and emission direction described by the spherical angles $\theta$ and $\phi$:
\begin{equation}\label{eq-4}
\tauWg(k, \theta, \phi) = \hbar\frac{\D~\mathrm{arg} \left(\sum_{l,m} \psi^\ast_{lm}(k,\theta,\phi) \right)}{\D~\epsilon}.
\end{equation} 
In the above equation, $\epsilon$ denotes the continuum electron energy, $\hbar$ is the reduced Planck constant and the quantity $\psi_{lm}$ denotes the partial waves in terms of which the outgoing wave function $\Psi_\mathrm{g}$ is expanded:
\begin{equation}
\Psi_\mathrm{g} = \sum_{lm}\psi_{lm}.
\end{equation}
We argue that the definition given in (\ref{eq-4}) is not consistent, neither with the established interpretation of time delay phenomena~\cite{Eisenbud48, Wigner55, Smith60, Nussenzveig72} nor the recent theoretical work on photoionization delays of atomic systems (see, e.g.~\cite{Dahlstroem13}). Following the derivation given by Wigner, the time delay $\tau$ in molecular photoionization can be related to the group delay of the outgoing photoelectron wave packet. This quantity is given by the energy derivative of the complex photoionization amplitude $f(\epsilon)$:
\begin{equation}
\tau = \frac{\D}{\D \epsilon} \mathrm{arg}\left( f(\epsilon)\right) = \mathrm{Im}\left\{ \frac{1}{f(\epsilon)}\frac{\D f}{\D \epsilon} \right\}.
\end{equation} 
A convenient practical route towards calculating $f(\epsilon)$ from first principles using the single-photon perturbation framework employed in \cite{Hockett16} is by performing a partial wave expansion in spherical waves:
\begin{equation}\label{fE}
f(\epsilon) = \sqrt{\frac{4\pi}{3}}\sum_{l,m}\bra{\psi_{lm}}r_\nu\ket{\psi_0} Y_{lm}(\Omega_{\hat{k}})Y_{1\nu}(\Omega_{\hat{\nu}}),
\end{equation} 
where the quantities $Y_{lm}$ denote the spherical harmonic functions describing the orientation of the outgoing photoelectron vector ($\Omega_{\hat{k}}=(\theta,\phi)$) and the photon po\-la\-ri\-za\-tion ($\Omega_{\hat{\nu}}$) directions. The exact form of the above equation may vary depending on the normalization conditions imposed on the continuum wave functions or the gauge (length vs. velocity). Differentiating the phase of $f(\epsilon)$ with respect to $\epsilon$, we obtain for the time delay $\tau$:
\begin{equation}\label{tau}
\tau(k,\theta,\phi,\Omega_{\hat{nu}}) = \hbar \frac{\D }{\D \epsilon} \mathrm{arg}\left( \sqrt{\frac{4\pi}{3}}\sum_{l,m}\bra{\psi_{lm}}r_\nu\ket{\psi_0} Y_{lm}(\Omega_{\hat{k}})Y_{1\nu}(\Omega_{\hat{nu}})\right) .
\end{equation}
 We emphasize that expression (\ref{tau}) contains two differences as compared to (\ref{eq-4}), namely the presence of dipole matrix elements $\bra{\psi_{lm}}r_\nu\ket{\psi_0}$ between the continuum waves and the initial state $\psi_0$ and the angular factors. Although the spatial contribution from the phase of the spherical harmonics is discussed in the original article, the dependence on the matrix elements has been neglected.\\
 
We note that the correct Equation (\ref{tau}) has been given and illustrated in our recent publications \cite{woerner15a,huppert16a,baykusheva16b}.

\end{document}